\RequirePackage[l2tabu,orthodox]{nag}
\documentclass
[11pt,letterpaper]
{article} 

\usepackage[notes=true,later=false,camera=true]{dtrt}
\usepackage[utf8]{inputenc}
\usepackage{etex}
\usepackage{ stmaryrd }
\usepackage{xspace,enumerate}
\usepackage[T1]{fontenc}
\usepackage[full]{textcomp}
\usepackage[american]{babel}
\usepackage{mathtools}

\usepackage{amsthm}
\usepackage{empheq}

   \usepackage{hyperref}
   \hypersetup{hyperindex=true,pdfpagemode=UseOutlines,bookmarksnumbered=true,bookmarksopen=true,bookmarksopenlevel=2,pdfstartview=FitH,pdfborder={0 0 1},linkbordercolor=dt@linkcolor,citebordercolor=dt@linkcolor,urlbordercolor=dt@linkcolor}
\usepackage[capitalise,nameinlink]{cleveref}
\crefname{lemma}{Lemma}{Lemmas}
\crefname{fact}{Fact}{Facts}
\newcommand{\colorconstraints}{\text{Color Constraints}}
\crefname{colorconstraints}{(color constraints)}{Color Constraints}
\crefformat{colorconstraints}{#2\colorconstraints#3}
\crefname{indsetconstraints}{(indset constraints)}{IndSet Constraints}
\crefformat{indsetconstraints}{#2$\mathsf{IndSet\ Axioms}$#3}
\crefname{theorem}{Theorem}{Theorems}
\crefname{mtheorem}{Theorem}{Theorems}
\crefname{corollary}{Corollary}{Corollaries}
\crefname{claim}{Claim}{Claims}
\crefname{example}{Example}{Examples}
\crefname{algorithm}{Algorithm}{Algorithms}
\crefname{problem}{Problem}{Problems}
\crefname{definition}{Definition}{Definitions}
\usepackage{paralist}
\usepackage{turnstile}
\usepackage{mdframed}
\usepackage{tikz}
\usepackage{caption}
\DeclareCaptionType{Algorithm}
\usepackage{newfloat}
\newtheorem{theorem}{Theorem}[section]
\newtheorem*{theorem*}{Theorem}

\newtheorem*{proposition*}{Proposition}
\newtheorem{lemma}[theorem]{Lemma}
\newtheorem*{lemma*}{Lemma}

\newtheorem*{conjecture*}{Conjecture}
\newtheorem{fact}[theorem]{Fact}
\newtheorem*{fact*}{Fact}

\newtheorem*{hypothesis*}{Hypothesis}

\theoremstyle{definition}

\newtheorem*{definition*}{Definition}

\theoremstyle{remark}
\newtheorem{claim}[theorem]{Claim}
\newtheorem*{claim*}{Claim}
\newtheorem{remark}[theorem]{Remark}
\newtheorem*{remark*}{Remark}

\newtheorem*{observation*}{Observation}

\usepackage[
letterpaper,
top=1.2in,
bottom=1.2in,
left=1in,
right=1in]{geometry}
\usepackage{newpxtext} 
\usepackage{textcomp} 
\usepackage[varg,bigdelims]{newpxmath}
\usepackage[scr=rsfso]{mathalfa}
\usepackage{bm} 
\let\mathbb\varmathbb
\usepackage{microtype}

\allowdisplaybreaks
\newcommand{\FormatAuthor}[3]{
\begin{tabular}{c}
#1 \\ {\small\texttt{#2}} \\ {\small #3}
\end{tabular}
}

\newcommand{\R}{{\mathbb R}}
\newcommand{\N}{{\mathbb N}}
\newcommand{\norm}[1]{\lVert #1 \rVert}

\newcommand{\abs}[1]{\lvert #1 \rvert}

\newcommand{\eps}{\varepsilon}

\newcommand{\E}{{\mathbb E}}

\newcommand{\cH}{\mathcal H}

\newcommand{\poly}{\mathrm{poly}}

\newcommand{\polylog}{\mathrm{polylog}}

\newcommand{\Bin}{\textrm{Bin}}
\newcommand{\cA}{\mathcal A}
\newcommand{\tO}{\tilde{O}}
\newcommand{\Lovasz}{Lov\'asz }
\newcommand{\alg}{\text{alg}}
\let\svthefootnote\thefootnote
\newcommand\blfootnote[1]{%
  \let\thefootnote\relax%
  \footnotetext{#1}%
  \let\thefootnote\svthefootnote%
}

\begin{document}

\title{Bypassing the XOR Trick: Stronger Certificates for Hypergraph Clique Number}
\author{
\begin{tabular}[h!]{ccc}
      \FormatAuthor{Venkatesan Guruswami\thanks{Supported in part by NSF grant CCF-1908125 and a Simons Investigator award.}}{venkatg@berkeley.edu}{UC Berkeley}
      \FormatAuthor{Pravesh K.\ Kothari\thanks{Supported in part by an NSF CAREER Award \#2047933, a Google Research Scholar Award, and a Sloan Fellowship.}}{praveshk@cs.cmu.edu }{Carnegie Mellon University}
            \FormatAuthor{Peter Manohar\thanks{Supported in part by an ARCS Scholarship, NSF Graduate Research Fellowship (under grant numbers DGE1745016 and DGE2140739), and NSF CCF-1814603.}}{pmanohar@cs.cmu.edu}{Carnegie Mellon University}
\end{tabular}
}
\date{}

\parskip=1ex

\maketitle\blfootnote{Any opinions, findings, and conclusions or recommendations expressed in this material are those of the author(s) and do not necessarily reflect the views of the National Science Foundation.}
	\thispagestyle{empty}

\begin{abstract}
Let $\cH(k,n,p)$ be the distribution on $k$-uniform hypergraphs where every subset of $[n]$ of size $k$ is included as an hyperedge with probability $p$ independently. In this work, we design and analyze a simple spectral algorithm that certifies a bound on the size of the largest clique, $\omega(H)$, in hypergraphs $H \sim \cH(k,n,p)$. For example, for any constant $p$, with high probability over the choice of the hypergraph, our spectral algorithm certifies a bound of $\tilde{O}(\sqrt{n})$ on the clique number in polynomial time. This matches, up to $\polylog(n)$ factors, the best known certificate for the clique number in random graphs, which is the special case of $k = 2$.

Prior to our work, the best known refutation algorithms~\cite{Coja-OghlanGL04,AOW15} rely on a reduction to the problem of refuting random $k$-XOR via Feige's XOR trick~\cite{Feige02}, and yield a polynomially worse bound of $\tilde{O}(n^{3/4})$ on the clique number when $p = O(1)$. Our algorithm bypasses the XOR trick and relies instead on a natural generalization of the \Lovasz theta semidefinite programming relaxation for cliques in hypergraphs. 
\end{abstract}


\clearpage
\section{Introduction}
In this work, we study the average-case search problem of finding refutations, i.e., certificates of the tightest possible upper bounds, on the clique number $\omega(H)$ (size of the largest clique) of random $k$-uniform hypergraphs $H$ drawn from the distribution $\cH(k, n, p)$, where each hyperedge is included in $H$ independently with probability $p$. 
A clique in a $k$-uniform hypergraph $H$ is a set $S$ of vertices such that all subsets $C \subseteq S$ with $\abs{C} = k$ are edges in the hypergraph $H$, and we will adopt the convention that if $\abs{S} \leq k-1$, then $S$ is trivially a clique.
With high probability, the clique number of such hypergraphs is $O(\log(n)^{\frac{1}{k-1}})$ for constant $p$, and approaches $n$ as $p$ approaches $1 - O(n^{-(k-1)})$ \cite{KrivelevichS98}.
Our goal is to find polynomial time algorithms that certify a bound as close to this ground truth as possible. 

In the case of Erdös-Renyi random graphs from $G(n, p)$, i.e., when $k = 2$, the \Lovasz theta function provides a semidefinite programming relaxation that certifies a bound of $O(\sqrt{n})$ in polynomial time with high probability over the draw of the graph, when $p = O(1)$.  A long line of work \cite{FeigeK00,FeigeK03,MekaPW15,DeshpandeM15,HopkinsKP15,RaghavendraS15,BarakHKKMP16} has explored the power of spectral methods and semidefinite programming hierarchies for improving on this bound. This question is also closely related to the \emph{planted clique} problem \cite{Jerrum92,Kucera95,AlonKS98,FeigeK00,FeigeK03}, where the size of the cliques that can be efficiently recovered is similar to the best known polynomially-certifiable upper bounds on the clique number of random graphs.

For $k>2$, the problem was first studied by Coja-Oghlan, Goerdt and Lanka~\cite{Coja-OghlanGL04}, who provided a polynomial time algorithm based on spectral methods to certify an upper bound of $\eps n$ on the clique number of random $3$- and $4$-uniform hypergraphs, where $\eps$ is a constant. Unlike the case of $k=2$, their algorithm relies on a reduction, via the famous XOR trick of Feige~\cite{Feige02}, to the problem of refuting random $k$-XOR formulas. Specifically, they construct a polynomial $f(x) = \frac{1}{{n \choose k}} \sum_{C \in {[n] \choose k}} b_C \prod_{i \in C} x_i$, where $b_C = 1-p$ if $C \in H$ and $b_C = -p$ if $C \notin H$, and they show that \begin{inparaenum}[(1)] \item if $\omega(H)$ is large, then $f(x)$ is large for some $x \in [-1,1]^n$, and \item with high probability over $H \sim \cH(k, n, p)$, their $k$-XOR refutation algorithm certifies a nontrivial upper bound on $\max_{x \in [-1,1]^n} f(x)$, and thus on $\omega(H)$\end{inparaenum}.
This connection was later utilized by Allen, O'Donnell and Witmer~\cite{AOW15} who used their $k$-XOR refutation algorithms to improve the bounds and handle the case of all $k \geq 3$. For any $k$, when $p = O(1)$, the algorithm of \cite{AOW15} certifies $\omega(H) \leq \tilde{O}(n^{3/4})$.

In this paper, we show that the certificates obtained via the ``XOR method'' are in fact suboptimal by providing a substantially improved certificate for the clique numbers of random hypergraphs at all densities $p$. Our certificates are based on a natural generalization of the ``direct'' \Lovasz SDP for clique numbers of graphs. The bounds obtained by our algorithm for any fixed $k$ match those obtained in the case of graphs (i.e., $k=2$) up to polylogarithmic factors in $n$.\footnote{We believe that with a more fine-grained analysis, our bound can be improved from $\tilde{O}(\sqrt{n})$ to $O(\sqrt{n})$. However, for simplicity we use ``off-the-shelf'' concentration inequalities, which lose $\polylog(n)$ factors over sharper methods.} Specifically, we show:
\begin{theorem}[\cref{thm:main}, specialized to $\poly(n)$-time and $p = O(1)$]\label{ithm:main}
There is an algorithm that takes as input a $k$-uniform hypergraph $H$ on $n$ vertices and a parameter $p \in [0,1]$ with $p = O(1)$, and outputs in $n^{O(k)}$-time a value $\omega_{\alg}(H) \in [0, n]$ with the following two properties:
\begin{enumerate}[(1)]
\item\label{item:completeness} Completeness: $\omega(H) \leq \omega_{\alg}(H)$, for all $H$.
\item\label{item:usefulness} High probability bound: If $H \sim \cH(k, n, p)$, then with probability $1 - 1/\poly(n)$, $\omega_{\alg}(H) \leq \tilde{O}(\sqrt{n})$.
\end{enumerate}
\end{theorem} 
The result above is based on a (surprisingly) simple spectral algorithm that is a natural analog, for $k$-uniform hypergraphs, of the algorithm that uses the spectral norm of the adjacency matrix of a graph to certify an upper bound on its clique number. This is in contrast to the methods from \cite{Coja-OghlanGL04,AOW15} that rely on a reduction to refuting random $k$-XOR formulas. 

It is easy to observe that a clique in $H$ is an independent set in the complement hypergraph $\overline{H} := \{C : C \notin H\}$, and that $\overline{H} \sim \cH(k,n,1-p)$ when $H \sim \cH(k,n,p)$. Thus, \cref{ithm:main} certifies a bound of $\tilde{O}(\sqrt{n})$ on the size of the maximum independent set in a random $H \sim \cH(k,n,1-p)$ with high probability, and hence also certifies with high probability that the chromatic number of a random $H$ is at least $\tilde{\Omega}(\sqrt{n})$.

\cref{ithm:main} is a special case of our more general theorem (\cref{thm:main}), which we present in full in \cref{sec:proof}. The algorithm in \cref{thm:main} has a tradeoff between the runtime and the strength of the certificate, and also handles the more general case of nonconstant $p$.

\begin{remark}[Detecting planted cliques vs.\ refutation]
We note that it is easy to \emph{distinguish} between a random $H \sim \cH(k,n,1/2)$ and a random $H \sim \cH(k,n,1/2)$ with a \emph{planted} clique of size $\tilde{\Theta}(\sqrt{n})$; the extra $\polylog(n)$ factor makes the degrees of vertices in the planted clique be noticeably larger than other vertices, so one can easily extract the planted clique. However, we are considering the formally harder task of \emph{refutation}, so merely being able to distinguish a random $H \sim \cH(k,n,1/2)$ from a $H$ from this particular planted distribution is insufficient. For example, it is easy to construct a hypergraph $H$ where the vertices in the planted clique do not have larger-than-average degree, which would, e.g., trivially fool the aforementioned simple distinguisher.
\end{remark}
\section{The Spectral Algorithm}
\label{sec:proof}
In this section, we prove the following theorem, which has \cref{ithm:main} as a special case. 
\begin{theorem}\label{thm:main}
There is an algorithm that takes as input a $k$-uniform hypergraph $H$ on $n$ vertices, a parameter $p := p(n) \in [0,1]$, and an integer $d := d(n) \geq 1$, and outputs in $n^{O(d + k)}$-time a value $\omega_{\alg}(H) \in [0, n]$ with the following two properties:
\begin{enumerate}[(1)]
\item\label{item:completeness} Completeness: $\omega(H) \leq \omega_{\alg}(H)$, for all $H$.
\item\label{item:usefulness} High probability bound: If $H \sim \cH(k, n,p)$, then with probability $1 - 1/\poly(n)$, 
$$\omega_{\alg}(H) \leq d + O(k) \left(\frac{d\log^2 n}{1-p}\right)^{\frac{2}{k'}} \sqrt{\max(n p^{{d \choose k' - 1}}, d \log n)} \ , $$ 
where $k' = k$ if $k$ is even, and $k' = k - 1$ if $k$ is odd.
\end{enumerate}
\end{theorem}
Before continuing with the proof, we first interpret \cref{thm:main} and compare it to the works of \cite{Coja-OghlanGL04,AOW15}.

When $k = 3$, \cite{Coja-OghlanGL04} certifies $\omega(H) \leq \eps n$ for constant $\eps$ when $p \leq 1 - O(n^{-3/2})$, and when $k = 4$, \cite{Coja-OghlanGL04} certifies $\omega(H) \leq \eps n$ for constant $\eps$ for $p \leq 1 - O(\frac{1}{n^2})$. \cite{AOW15} improves upon \cite{Coja-OghlanGL04} and certifies $\omega(H) \leq \tO(n^{3/4 + \theta/2k})$ when $p = 1 - \tO(n^{-\theta})$; in particular, for, e.g., $p = \frac{1}{2}$, they certify that $\omega(H) \leq \tO(n^{3/4})$, and this bound gets worse as $p$ increases. When $p = O(1)$, our algorithm certifies a bound of $\omega \leq \tO(\sqrt{n})$ in polynomial time for any fixed $k$. \cref{thm:main} thus beats the current best known algorithm in \cite{AOW15} by a factor of $n^{1/4}$, i.e., a \emph{polynomial} factor.

 More generally, for any $d \in \N$, our algorithm certifies a bound of $\omega \leq \tO(1) + \tO(\sqrt{n p^{d \choose k' - 1}})$ in time $n^{O(d + k)}$. On the other hand, for $H \sim \cH(k,n,p)$ with $p \leq 1 - O(n^{-(k-1)})$, \cite{KrivelevichS98} shows that with high probability, the true clique size is $\omega(H) \leq O((1-p)^{-\frac{1}{k-1}} (\log(n^{k-1}(1-p)))^{\frac{1}{k-1}})$. So, for, e.g., $d = O(\log n)$ and $p \leq O(1)$, our algorithm outputs $\omega_{\alg}(H) \leq \tO(1)$, which is the true clique number up to $\polylog(n)$ factors.

We now turn to the proof of \cref{thm:main}.
\begin{proof}[Proof of \cref{thm:main}]
We break the proof of \cref{thm:main} into three steps. First, we give a simple refutation algorithm that achieves the guarantees of \cref{thm:main} when $k$ is even and $d = 1$. Then, we prove the case when $k$ is even and $d$ is arbitrary by reduction to the case when $d = 1$. Finally, we reduce the case of odd $k$ to the case of even $k$.

For a set $S$ and integer $t$, we will let ${S \choose t}$ denote the set of all subsets of $S$ of size exactly $t$. E.g., ${[n] \choose k} := \{C \subseteq [n], \abs{C} = k\}$.

\parhead{The basic refutation algorithm.}
We first give a basic refutation algorithm. This algorithm achieves the guarantees of \cref{thm:main} in the case when $k$ is even and $d = 1$.
\begin{lemma}\label{lem:polytimealg}
Let $k$ be even. There is an algorithm $\cA$ that takes as input a $k$-uniform hypergraph $H$ on $n$ vertices and a parameter $p := p(n) \in [0,1]$, and outputs in $n^{O(k)}$-time a value $\omega_{\alg}(H) \in [0, n]$ with the following two properties:
\begin{enumerate}[(1)]
\item\label{item:polytimecompleteness} Completeness: $\omega(H) \leq \omega_{\alg}(H)$, for all $H \subseteq {[n] \choose k}$.
\item\label{item:polytimeusefulness} High probability bound: If $H \sim \cH(k, n,p)$, then for any $c \geq 1$, with probability $1 - n^{-c}$, $$\omega_{\alg}(H) \leq O(k) \left(\frac{c \log n}{1-p}\right)^{\frac{2}{k}} \sqrt{n} \enspace. $$
\end{enumerate}
\end{lemma}
 We prove \cref{lem:polytimealg} in \cref{sec:polytimealg}.

\parhead{Case 1: $k$ is even.}
We now prove \cref{thm:main} when $k$ is even. To do this, we need the following claim.
\begin{claim}\label{claim:dreduction}
Let $H$ be a $k$-uniform hypergraph on $n$ vertices, let $d \geq k-1$ be a positive integer, and let $J \subseteq [n]$ be a set of size $d$. Let $V_J = \{i \in [n] \setminus J : \forall J' \in {J \choose k-1}, J' \cup \{i\} \in H\}$, and let $H_J = {V_J \choose k} \cap H$. Then, $\omega(H) \leq d + \max_{J \in {n \choose d}} \omega(H_J)$. Moreover, if $H \sim \cH(k, n, p)$, then $\abs{V_J} \sim \Bin(n - d, p^{{d \choose k-1}})$, and conditioned on $V_J$, $H_J \sim \cH(k, |V_J|, p)$.
\end{claim}
We prove \cref{claim:dreduction} in \cref{sec:dreduction}

With \cref{claim:dreduction} in hand, we finish the proof of \cref{thm:main} when $k$ is even. Let $\cA$ be the algorithm in \cref{lem:polytimealg}, and let the algorithm $\cA'$ operate as follows: on input $H$, \begin{inparaenum}[(1)] \item enumerate over all $J \in {[n] \choose d}$ and compute $\cA(H_J)$, and then \item output $d + \max_{J \in {[n] \choose d}} \cA(H_J)$\end{inparaenum}. Clearly, $\cA'$ runs in $n^{O(d)} \cdot n^{O(k)} = n^{O(d + k)}$ time. We observe that by \cref{lem:polytimealg,claim:dreduction}, we clearly have that $\omega(H) \leq d + \max_{J \in {[n] \choose d}} \omega(H_J) \leq d + \max_{J \in {[n] \choose d}} \cA(H_J) = \cA'(H)$, so \cref{item:completeness} holds. 
We now prove \cref{item:usefulness}. Fix $J \in {[n] \choose d}$. We observe that by \cref{claim:dreduction}, $V_J \sim \Bin(n - d, p^{ {d \choose k-1}})$. We bound $\abs{V_J}$ using the standard Chernoff bound, which we recall below.
\begin{fact}[Chernoff Bound]\label{fact:chernoff}
Let $X \sim \Bin(n,p)$, and let $\delta \geq 0$. Then, $\Pr[X \geq (1 + \delta)np] \leq \exp(\frac{-\delta^2 n p}{2 + \delta})$.
\end{fact}
\cref{fact:chernoff} implies that $\abs{V_J} \leq \max(n p^{{d \choose k - 1}}, d \log n)$ with probability $\geq 1 - n^{-2d}$. Now, conditioned on $\abs{V_J}$, by \cref{claim:dreduction} we have that $H_J \sim \cH(k, \abs{V_J}, p)$. Hence, if $\abs{V_J} \geq 1$, setting $c = O(d \log n)$ in \cref{lem:polytimealg}, we have that with probability $\geq 1 - n^{-2d}$, $\cA(H_J) \leq O(k) \left(\frac{d \log^2 n}{1-p}\right)^{\frac{2}{k}} \sqrt{\abs{V_J}}$. (If $\abs{V_J} = 0$, then $\cA(H_J) = k - 1$.) Hence, for a fixed $J$, with probability $\geq 1 - 2n^{-2d}$, we have $\cA(H_J) \leq O(k) \left(\frac{ d \log^2 n}{1-p}\right)^{\frac{2}{k}} \sqrt{\max(n p^{{d \choose k - 1}}, d \log n)}$. By union bound over all $J$, we thus conclude that $\cA'(H) \leq d + O(k) \left(\frac{d\log^2 n}{1-p}\right)^{\frac{2}{k}} \sqrt{\max(n p^{{d \choose k - 1}}, d \log n)}$ with probability $\geq 1 - 2n^{-d} = 1 - 1/\poly(n)$. This finishes the proof of \cref{thm:main} when $k$ is even.

\parhead{Case 2: $k$ is odd}
We now turn to the case when $k$ is odd. For the odd case, we use the following claim, which we prove in \cref{sec:oddreduction}.
\begin{claim}\label{claim:oddreduction}
Let $H$ be a $k$-uniform hypergraph on $n$ vertices with $k \geq 3$. For each $i \in [n]$, let $H_i = \{C \setminus \{i\} : C \in H \wedge i \in C\}$. Then, $\omega(H) \leq 1 + \max_{i \in [n]} \omega(H_i)$. Moreover, if $H \sim \cH(k, n, p)$, then for any fixed $i \in [n]$, $H_i \sim \cH(k, n-1, p)$. 
\end{claim}
Let $\cA'$ be the algorithm in \cref{thm:main} when $k$ is even, described earlier. Let $\cA''$ be the algorithm that operates as follows: on input $H$, a $k$-uniform hypergraph where $k$ is odd, \begin{inparaenum}[(1)]
\item for each $i \in [n]$, compute $\cA'(H_i)$, \item output $1 + \max_{i \in [n]} \cA'(H_i)$\end{inparaenum}. Clearly, $\cA''$ runs in $n^{O(d + k)}$ time, and by \cref{claim:oddreduction}, we have that $\omega(H) \leq 1 + \max_{i \in [n]} \omega(H_i) \leq 1 + \max_{i \in [n]} \cA'(H_i) = \cA''(H)$. Thus, \cref{item:completeness} holds. To see \cref{item:usefulness}, we observe that by \cref{claim:oddreduction}, $H_i \sim \cH(k, n-1, p)$. Hence, with probability $\geq 1 - 2n^{-d}$, it holds that $\cA'(H_i) \leq d + O(k) \left(\frac{d \log^2 n}{1-p}\right)^{\frac{2}{k-1}} \sqrt{\max(n p^{{d \choose k - 2}}, d \log^2 n)}$. By union bound over the choice of $i$, we see that with probability $1 - 1/\poly(n)$, 
$$ \cA''(H) \leq d + O(k) \left(\frac{d \log^2 n}{1-p}\right)^{\frac{2}{k-1}} \sqrt{\max(n p^{{d \choose k - 2}}, d \log n)} \ , $$
which finishes the proof of \cref{thm:main}.
\end{proof}
\subsection{The basic algorithm: proof of \cref{lem:polytimealg}}
\label{sec:polytimealg}
\begin{proof}
For $C \in {[n] \choose k}$, let $A_C \in \R^{{[n] \choose k/2} \times {[n] \choose k/2}}$ be the matrix where $A_C(S,T) = 1$ if $S \cup T = C$, and $0$ otherwise. Note that this implies that $S \cap T = \emptyset$ also.

Let $A = \sum_{C \in {[n] \choose k}} b_C A_C$, where $b_C = 1 - p$ if $C \in H$, and $b_C = -p$ if $C \notin H$. The output of the algorithm is $\omega_{\alg}(H) := \omega = \max(k-1, k \left(\frac{p}{1-p} {k \choose k/2} \cdot \norm{A}_2^2\right)^{\frac{1}{k}})$. We observe that $A$ can be constructed in $n^{O(k)}$ time and has size $n^{O(k)}$, and so we can compute $\norm{A}_2$ (and thus also $\omega$) in $n^{O(k)}$ time. 

We now prove \cref{item:polytimecompleteness}. Let $I \subseteq [n]$ be a clique in $H$. If $\abs{I} \leq k-1$, then we are done, as $\omega \geq k - 1$ always holds. So, suppose that $\abs{I} \geq k$. Let $x \in \R^{{[n] \choose k/2}}$ be defined as $x_S = 1$ if $S \subseteq I$, and $0$ otherwise. Note that $\norm{x}_2^2 = {\abs{I} \choose k/2}$. We observe that $x^{\top} A x$ is simply $\sqrt{\frac{1-p}{p}} {\abs{I} \choose k} \cdot {k \choose k/2}$. This is because for each $C \in {I \choose k}$, there are ${k \choose k/2}$ ways to partition $C$ into $(S,T)$, and all such $C$ are in $H$, and thus $b_C = 1-p$. We thus conclude that 
\begin{flalign*}
&\norm{A}_2 \geq \frac{x^{\top} A x}{\norm{x}_2^2} = (1-p) {\abs{I} \choose k} \cdot {k \choose k/2} \cdot \frac{1}{{\abs{I} \choose k/2}} \geq (1-p)\sqrt{{\abs{I} \choose k}  {k \choose k/2}} \geq (1-p)\sqrt{\left(\frac{\abs{I}}{k}\right)^k  {k \choose k/2}} \enspace,
\end{flalign*}
where we use that $\frac{{\abs{I} \choose k}}{{\abs{I} \choose k/2}^2} \geq \frac{1}{{k \choose k/2}}$, and that ${\abs{I} \choose k} \geq \left(\frac{\abs{I}}{k}\right)^k$. Hence, we have shown that
\begin{flalign*}
\omega \geq k \left(\frac{1}{(1-p)^2} {k \choose k/2} \cdot \norm{A}_2^2\right)^{\frac{1}{k}} \geq \abs{I}
\end{flalign*}
for any clique $I$ in $H$ with $\abs{I} \geq k$. This finishes the proof of \cref{item:completeness}.

We now prove \cref{item:polytimeusefulness}. The key step here is to show an upper bound on $\norm{A}_2$, with high probability over $H \sim \cH(k, n,p)$.
We will do this by applying the standard Matrix Bernstein concentration inequality, which we recall below.
\begin{fact}[Matrix Bernstein, Theorem 1.4 of \cite{Tropp2012}]
\label{fact:matrixbernstein}
Let $X_1, \dots, X_k$ be independent random $n \times n$ symmetric matrices with $\E[X_i] = 0$ and $\norm{X_i}_2 \leq R$ for all $i$. Let $\sigma^2 \geq \norm{\E[\sum_{i = 1}^k X_i^2]}_2$. Then for any $c > 0$, $\Pr[\norm{\sum_{i = 1}^k X_i}_2 \geq O(Rc\log n + \sigma \sqrt{c\log n})] \leq n^{-c}$.
\end{fact}
We observe that $A = \sum_{C \in {[n] \choose k}} b_C A_C$ is the sum of ${n \choose k}$ independent, mean $0$ random matrices.
We have that $\norm{A_C}_2 \leq R := \max(p, 1-p) = 1$ for every $C$, as each row/column of $A_C$ has at most one nonzero entry, which is at most $R$ in magnitude. We also observe that $\E[A^2] = \sum_{C} A_C^2$ is a diagonal matrix, where the $S$-th diagonal entry is ${n - k/2 \choose k - k/2} = {n - k/2 \choose k/2} \leq n^{k/2}$, as each $A_C^2$ is diagonal and has a $1$ in the $S$-th diagonal entry if $S \subseteq C$, as $\E[b_C^2] = 1$.
Hence, by \cref{fact:matrixbernstein}, with probability $1 - n^{-c}$, we have that 
\begin{flalign*}
&\norm{A}_2 \leq O(Rc\log n^{k/2}) + O(\sqrt{c n^{k/2} \log n^{k/2}}) \leq O(n^{\frac{k}{4}} c k \log n) \\
&\implies k \left(\frac{1}{(1-p)^2} {k \choose k/2} \cdot \norm{A}_2^2\right)^{\frac{1}{k}} \leq k \left(\frac{1}{(1-p)^2} {k \choose k/2} \cdot O(n^{k/2} c^2 k^2 \log^2 n)\right)^{\frac{1}{k}} \leq O(k) \left(\frac{ c \log n}{1-p}\right)^{\frac{2}{k}} \sqrt{n} \enspace.
\end{flalign*}
Finally, we have that
\begin{flalign*}
\omega = \max(k-1, O(k) \left(\frac{ c \log n}{1-p}\right)^{\frac{2}{k}} \sqrt{n}) = O(k) \left(\frac{ c \log n}{1-p}\right)^{\frac{2}{k}} \sqrt{n} \enspace. {\hfill \qedhere}
\end{flalign*}
\end{proof}

\subsection{Reduction for larger $d$: proof of \cref{claim:dreduction}}
\label{sec:dreduction}

\begin{proof}
Let $I$ be a clique in $H$ with $\abs{I} = \omega(H)$. Let $J \subseteq I$ be an arbitrary subset of size $d$. We claim that $I' := I \setminus J$ is a clique in $H_J$. Indeed, we first observe that for each $i \in I'$, we have $i \in V_J$, as for any $J' \in {J \choose k-1}$, we have $J' \cup \{i\}$ is a subset of $I$ of size $k$, and hence is in $H$. Next, let $C \subseteq I'$ be any subset of size $k$ (if $\abs{I'} \leq k - 1$, so that no such $C$ exists, then $I'$ is trivially a clique in $H_J$). Then, $C \in H$, as $I$ was a clique, and so $C \in H_J$. Hence, $I'$ is a clique in $H_J$. As $\omega(H) =  \abs{J} + \abs{I'} = d + \abs{I'} \leq d + \omega(H_J)$, this proves the first part of the claim.

For the second part of the claim, we think of sampling $H$ as follows. First, for every $J' \in {J \choose k-1}$ and $i \in [n]$, add $J' \cup \{i\}$ to $H$ with probability $p$. Then, add every other $C \in {[n] \choose k}$ to $H$ with probability $p$. We note that $H \sim \cH(k,n,p)$ clearly, and that after the first step, we have determined $V_J$. In the first step, we see that $i$ is added to $V_J$ independently for each $i \notin J$, and each $i$ is added with probability $p^{{d \choose k-1}}$. Hence, $\abs{V_J} \sim \Bin(n - d, p^{{d \choose k-1}})$. As all the hyperedges in $H_J$ are sampled in the second step, the claim follows.
\end{proof}
\subsection{Reduction from odd $k$ to even $k$: proof of \cref{claim:oddreduction}}
\label{sec:oddreduction}

\begin{proof}
Let $I \in H$ be a clique with $\abs{I} = \omega(H)$. If $\abs{I} = k - 1$, then we are done, as $\omega(H_i) \geq k - 2$ for all $i$ since $k \geq 3$. So, suppose $\abs{I} \geq k$. Let $i \in I$, and let $J = I \setminus \{i\}$. Then, $J$ is a clique in $H_i$. Indeed, for any $C' \in {J \choose k - 1}$, we must have $C' \cup \{i\} \in H$, and therefore we have $C' \in H_i$. So, it follows that $\omega(H_i) \geq \abs{J} = \abs{I} - 1$, which finishes the proof of the first part of the claim.

For the second part, we observe that if $H \sim \cH(k,n,p)$, then each $C \in {[n] \choose k}$ with $i \in C$ is added to $H$ independently with probability $p$. So, each $C' \in {[n] \setminus \{i\} \choose k - 1}$ is added to $H_i$ independently with probability $p$, which finishes the proof.
\end{proof}

\bibliographystyle{alpha}
\bibliography{hypergraph-indset.bbl}

\newcommand{\etalchar}[1]{$^{#1}$}
\begin{thebibliography}{BHK{\etalchar{+}}16}

\bibitem[AKS98]{AlonKS98}
Noga Alon, Michael Krivelevich, and Benny Sudakov.
\newblock Finding a large hidden clique in a random graph.
\newblock In {\em {SODA}}, pages 594--598. {ACM/SIAM}, 1998.

\bibitem[AOW15]{AOW15}
Sarah~R. Allen, Ryan O'Donnell, and David Witmer.
\newblock How to refute a random {CSP}.
\newblock In {\em Proceedings of the 56th Annual IEEE Symposium on Foundations
  of Computer Science}, pages 689--708, 2015.

\bibitem[BHK{\etalchar{+}}16]{BarakHKKMP16}
Boaz Barak, Samuel~B. Hopkins, Jonathan~A. Kelner, Pravesh Kothari, Ankur
  Moitra, and Aaron Potechin.
\newblock A nearly tight sum-of-squares lower bound for the planted clique
  problem.
\newblock In {\em {FOCS}}, pages 428--437. {IEEE} Computer Society, 2016.

\bibitem[CGL04]{Coja-OghlanGL04}
Amin Coja{-}Oghlan, Andreas Goerdt, and Andr{\'{e}} Lanka.
\newblock Strong refutation heuristics for random k-sat.
\newblock In {\em Approximation, Randomization, and Combinatorial Optimization,
  Algorithms and Techniques}, volume 3122 of {\em Lecture Notes in Computer
  Science}, pages 310--321. Springer, 2004.

\bibitem[DM15]{DeshpandeM15}
Yash Deshpande and Andrea Montanari.
\newblock Improved sum-of-squares lower bounds for hidden clique and hidden
  submatrix problems.
\newblock In {\em {COLT}}, volume~40 of {\em {JMLR} Workshop and Conference
  Proceedings}, pages 523--562. JMLR.org, 2015.

\bibitem[Fei02]{Feige02}
Uriel Feige.
\newblock Relations between average case complexity and approximation
  complexity.
\newblock In {\em {STOC}}, pages 534--543. {ACM}, 2002.

\bibitem[FK00]{FeigeK00}
Uriel Feige and Robert Krauthgamer.
\newblock Finding and certifying a large hidden clique in a semirandom graph.
\newblock {\em Random Struct. Algorithms}, 16(2):195--208, 2000.

\bibitem[FK03]{FeigeK03}
Uriel Feige and Robert Krauthgamer.
\newblock The probable value of the lov{\'{a}}sz--schrijver relaxations for
  maximum independent set.
\newblock {\em {SIAM} J. Comput.}, 32(2):345--370, 2003.

\bibitem[HKP15]{HopkinsKP15}
Samuel~B. Hopkins, Pravesh~K. Kothari, and Aaron Potechin.
\newblock Sos and planted clique: Tight analysis of {MPW} moments at all
  degrees and an optimal lower bound at degree four.
\newblock {\em CoRR}, abs/1507.05230, 2015.

\bibitem[Jer92]{Jerrum92}
Mark Jerrum.
\newblock Large cliques elude the metropolis process.
\newblock {\em Random Struct. Algorithms}, 3(4):347--360, 1992.

\bibitem[KS98]{KrivelevichS98}
Michael Krivelevich and Benny Sudakov.
\newblock The chromatic numbers of random hypergraphs.
\newblock {\em Random Struct. Algorithms}, 12(4):381--403, 1998.

\bibitem[Kuc95]{Kucera95}
Ludek Kucera.
\newblock Expected complexity of graph partitioning problems.
\newblock {\em Discrete Applied Mathematics}, 57(2-3):193--212, 1995.

\bibitem[MPW15]{MekaPW15}
Raghu Meka, Aaron Potechin, and Avi Wigderson.
\newblock Sum-of-squares lower bounds for planted clique.
\newblock In {\em {STOC}}, pages 87--96. {ACM}, 2015.

\bibitem[RS15]{RaghavendraS15}
Prasad Raghavendra and Tselil Schramm.
\newblock Tight lower bounds for planted clique in the degree-4 {SOS} program.
\newblock {\em CoRR}, abs/1507.05136, 2015.

\bibitem[Tro12]{Tropp2012}
Joel~A. Tropp.
\newblock User-friendly tail bounds for sums of random matrices.
\newblock {\em Foundations of Computational Mathematics}, 12(4):389--434, Aug
  2012.

\end{thebibliography}
\end{document}